\documentclass[]{aa}

\newcommand{\Mgas}{$M_{\rm gas}$}
\newcommand{\Mdust}{$M_{\rm dust}$}
\newcommand{\Mstar}{$M_{\rm \ast}$}
\newcommand{\Msol}{$\rm M_{\rm \odot}$}
\newcommand{\fgas}{$f_{\rm gas}$}
\newcommand{\fdust}{$f_{\rm dust}$}
\newcommand{\Gobat}{{\color{blue} G18}}
\newcommand{\Magdis}{{\color{blue} M21}}
\usepackage[dvipsnames]{xcolor}
\usepackage[export]{adjustbox}
\usepackage{graphicx}
\usepackage{txfonts}
\usepackage{gensymb}
\usepackage{tabularx}
\usepackage{multirow}
\usepackage{natbib,twoopt}
\usepackage[breaklinks=true]{hyperref}
\DeclareUnicodeCharacter{2212}{-}
\hypersetup{
    colorlinks=true,
    citecolor=blue,
    linkcolor=blue,
    filecolor=blue,      
    urlcolor=blue,
    }
\bibpunct{(}{)}{;}{a}{}{,}
\begin{document}

   \title{The gas mass reservoir of quiescent galaxies at cosmic noon}

   \author{David Bl\'anquez-Ses\'e\inst{1,2}
          \and
          C. Gómez-Guijarro\inst{3}
          \and
          G. E. Magdis\inst{1,2,4}%\fnmsep\thanks{Just to show the usage
          %of the elements in the author field}
          \and
          B. Magnelli \inst{3}
          \and
          R. Gobat\inst{5}
          \and
          E. Daddi\inst{3}
          \and
          M. Franco\inst{6}
          \and
          K. Whitaker\inst{7,1}
          \and
          F. Valentino\inst{8,1}
          \and
          S. Adscheid\inst{9}
          \and
          E. Schinnerer\inst{10}
          \and
          A. Zanella\inst{11}
          \and 
          M. Xiao\inst{12}
          \and 
          T. Wang\inst{13}
          \and
          D. Liu\inst{14}
          \and
          V. Kokorev\inst{1,15}
          \and
          D. Elbaz\inst{3}
          }
          
   \institute{ Cosmic Dawn Center (DAWN), Rådmandsgade 62, 2200 København, Denmark
   \and
            DTU Space, Technical University of Denmark, Elektrovej 327, DK-2800 Kgs. Lyngby,Denmark
   \and
             Universit\'e Paris-Saclay, Universit\'e Paris Cit\'e, CEA, CNRS, AIM, 91191, Gif-sur-Yvette, France
   \and
            Niels Bohr Institute, University of Copenhagen, Jagtvej 128, DK-2200 Copenhagen, Denmark
   \and 
            Instituto de Física, Pontificia Universidad Católica de Valparaíso, Casilla 4059, Valparaíso, Chile
   \and  
            Department of Astronomy, The University of Texas at Austin, 2515 Speedway Blvd Stop C1400, Austin, TX 78712, USA
   \and
            Department of Astronomy, University of Massachusetts, Amherst, MA 01003, USA
   \and
            European Southern Observatory, Karl-Schwarzschild-Str. 2, D-85748 Garching bei Munchen, Germany
   \and
            Argelander-Institut für Astronomie, Universit¨at Bonn, Auf dem H¨ugel 71, 53121 Bonn, Germany
   \and 
            Max-Planck-Institut für Astronomie, K’onigstuhl 17, D-69117, Heidelberg, Germany
   \and 
            Istituto Nazionale di Astrofisica (INAF), Vicolo dell’Osservatorio 5, I-35122 Padova, Italy
   \and 
            Department of Astronomy, University of Geneva, Chemin Pegasi 51, 1290 Versoix, Switzerland
   \and
            Key Laboratory of Modern Astronomy and Astrophysics (Nanjing
            University), Ministry of Education, Nanjing 210093, China
   \and
            Max-Planck-Institut für Extraterrestrische Physik (MPE), Giessenbachstraße 1, D-85748 Garching, Germany
   \and     
            Kapteyn Astronomical Institute, University of Groningen, P.O. Box 800, 9700AV Groningen, The Netherlands
             }

% \abstract{}{}{}{}{} 
% 5 {} token are mandatory
 
  \abstract
  {We present a 1.1mm stacking analysis of moderately massive (log(\Mstar/\Msol) = 10.7$\pm$0.2) quiescent galaxies (QGs) at $\langle z\rangle \sim1.5$, searching for cold dust continuum emission, an excellent tracer of dust and gas mass. Using both the recent GOODS-ALMA survey as well as the full suite of ALMA Band-6 ancillary data in the GOODS-S field, we report the tentative detection of dust continuum equivalent of   dust mass log(\Mdust/\Msol) = 7.47\,$\pm$\,0.13 and gas mass log(\Mgas/\Msol) = 9.42\,$\pm$\,0.14. The emerging gas fraction is \fgas\ = 5.3\,$\pm$\,1.8\%, consistent with the results of previous stacking analyses based on lower resolution sub(mm) observations. Our results support the scenario where high$-z$ QGs have an order of magnitude larger \fgas\ compared to their local counterparts and have experienced quenching with a non negligible gas reservoir in their interstellar medium - i.e. with gas retention. Subsequent analysis yields an anti-correlation between the  \fgas\ and the stellar mass of QGs, especially in the high mass end where galaxies reside in the most massive haloes. The \fgas\ - \Mstar\ anti-correlation promotes the selection bias as a possible solution to the tension between the stacking results pointing towards gas retention in high$-z$ QGs of moderate \Mstar\ and the studies of individual targets that favour a fully depleted ISM in massive (log(\Mstar/\Msol) > 11.2) high$-z$ QGs.}
 

   \maketitle 
   
%%%%%%%%%%%%%%%%%%%%%%%%%%%
%%%%%%%%%%%%%%%%%%%%%%%%%%%

\section{Introduction}
Understanding why galaxies die, i.e. become quiescent, is one of the most sought milestones in the current study of galaxy evolution. The existence of a population of dead galaxies, in contrast to the main sequence (MS) of star-forming galaxies (SFGs), has been confirmed and identified both photometrically \citep[e.g.][]{Daddi_2005,Daddi_2007,Toft_2005,Kriek_2006} and spectroscopically \citep{Toft_2012,Whitaker_2013,DEugenio_2020} up to \textit{z}\,$\sim$\,4 \citep{Valentino_2020} and even beyond \citep{Carnall_2022}. Quenched galaxies are characterized by very low levels or even total absence of star formation activity and red colours due to the old age of their evolved stellar populations. Their total comoving mass density has remarkably increased since \textit{z}\,$\sim$\,2.5 \citep{Tomczak_2014}, an epoch where the most massive galaxies (log(\Mstar/\Msol)\,$>$\,10.7) have seen their star formation rate (SFR) drastically reduced \citep{Muzzin_2013,Davidzon_2017}, signifying the onset of the decline of the star formation rate density of the Universe \citep{Madau_Dickinson_2014}.

The physical mechanisms responsible for the truncation of star formation in galaxies still remain uncertain, though many competing scenarios have been proposed, including: morphological quenching \citep{Martig_2009,Cornuault_2018}, active galactic nuclei (AGN) \citep{Di_Matteo_2005} and stellar feedback \citep{Ciotti_1991}, virial shock heating due to massive halos \citep{Rees_1977}, or cosmological starvation \citep{Feldmann_2015} (see \cite{Man_Belli} for further information).

The critical role of the interstellar medium (ISM) and more importantly of the cold molecular gas in the regulation of star formation and therefore the quenching of galaxies has also been established, playing a crucial part in all the proposed mechanisms. Measuring the amount of gas left in the ISM of quiescent galaxies (QGs) is pivotal to progress in this field. Furthermore, it is essential to observe these systems at high-\textit{z} where they offer an opportunity to study QGs closer to the quenching episode, making them prime laboratories to test the various quenching processes.

The molecular gas mass (\Mgas) and subsequently the gas fraction (\fgas\ = \Mgas/\Mstar) of local QGs has been measured and found to range between 0.3-1\,\%, typically $\sim\,100$ times lower than that of local SFGs \citep{Young_2011,Cappellari_2013,Boselli_2014,Davis_2014,Lianou_2016}. Quite naturally, the detection of the observable signatures of such low levels of gas reservoirs is becoming progressively more difficult as we move to higher redshifts, posing an observational challenge for the study of the ISM of distant QGs. Nevertheless, there have been several attempts to measure the gas reservoir in high-\textit{z} QGs and trace the evolution of their gas fraction with redshift, utilising CO \citep{Sargent_2015,Suess_2017,Hayashi_2018,Spilker_2018,Bezanson_2019,Belli_2021,Williams_2021}, [CII] \citep{Schreiber_2018_Jekyll} and [CI] \citep{Suzuki_2022} line observations of individual galaxies as well as dust continuum emission of stacked ensembles \citep{Gobat_2018,Magdis_2021} (hereafter \Gobat\ and \Magdis) or gravitationally lensed systems \citep{Caliendo_2021,Whitaker_2021}.

Intriguingly, these nascent studies have reached inconclusive and somewhat contradicting results. For example, CO(2-1) observations for a hand-full of very massive (log(\Mstar/$M_{\rm \odot})\,>\,11.55$) QGs at $1.2\,<\,z\,<\,1.5$ point towards negligible \fgas\ $<$ 1\,\% remaining gas reservoirs in their ISM \citep{Williams_2021}. A similar result was found with dust continuum observations of $1 < z < 3$ lensed QGs \citep{Whitaker_2021} obtained with the Atacama Large Millimiter/submillimeter Array (ALMA). Despite the aid of lensing amplification, the majority of the targets remained undetected at 1.1\,mm placing stringent upper limits on their \fgas\ comparable to that of local QGs (although see \Gobat). On the other hand, far-infrared/millimeter (FIR/mm) stacking analysis of massive (log($M_{\rm *}/M_{\odot})\,>\,10.7$), $0.5\,<\,z\,<\,2.5$ QGs (\Gobat\ , \Magdis) recovered dust continuum emission indicative of the presence of substantial amounts of gas and relatively large gas fractions, \fgas\ $\sim 5-10\%$, i.e. $\sim\times\,$50 larger than that of local ellipticals. 
 
Although these two contradicting results, fully depleted versus partially depleted ISM in distant QGs, possibly point towards a range of quenching mechanisms, any attempt to draw robust conclusions is hampered by selection effects and systematic uncertainties in the conversion of the observables (i.e., line luminosities, dust continuum emission) to \Mgas. Spectroscopic observations of the gas tracers (CO, [CII], [CI]) of high-\textit{z} QGs are still scarce and focused exclusively on the very massive end of the population while the analysis of gravitationally lensed systems might also suffer from surface brightness limitations, small sample sizes and other caveats which can introduce systematic uncertainties \citep{Gobat_2022}. On the other hand, the coarse resolution of the FIR/mm observations employed in the stacking analysis of \Gobat\ and \Magdis\ (\textit{Herschel}, JCMT/SCUBA-2, ASTE/AzTEC) can result in blending and contamination of the recovered signal and, thus, of the inferred \Mgas\ (and \fgas). The tension between the two approaches, stacking vs individual galaxies, could also stem from selection bias, with individual observations targeting the most exotic, massive high-\textit{z} QGs, while stacking being representative of the average, less massive QGs. 

In this work we aim to push the field a step further and overcome the main caveat of the stacking results presented in \Gobat\ and \Magdis, i.e the poor resolution of the FIR/mm observations. Namely, we present a stacking analysis of $1 < z < 3$ QGs by exploiting high angular resolution ALMA observations at 1.1\,mm and we present new measurements of the \Mgas\ and \fgas\ of the population. In Sect.~2 and 3 we present the data and the sample selection of our study. In Sect.~4 we describe the stacking analysis and the recovered signal. In Sect.~5 we present the measured dust mass (\Mdust) and resulting \Mgas\ and trace the evolution of \fgas\ with redshift, while in Sect.~6 we discuss the implication of our results and explore a potential \fgas\ $-$ \Mstar\ anti-correlation for high-\textit{z} QGs. In Sect.~7 we summarise our conclusions.

Throughout this work, we assume a standard $\Lambda$CDM cosmology with $\Omega_{M}$ = 0.3, $\Omega_{\Lambda}$ = 0.7 and $H_{0}$ = 70 km s$^{-1}$ Mpc$^{-1}$, adopt the Salpeter initial mass function (IMF) \citep{Salpeter_1955} and the AB magnitude system.

%%%%%%%%%%%%%%%%%%%%%%%%%%%
%%%%%%%%%%%%%
%%%%%%%%%%%%%%%%%%%%%%%%%%%
\section{ALMA data}
In this work we focus on QGs lying in GOODS-S \citep{GOODS_S} taking advantage of the rich multi-wavelength ancillary data in the field that we combine with 1.1\,mm data from the GOODS-ALMA survey \citep{Franco_2018,Gomez_2021}. As a second step, in order to maximise the sensitivity of our stacking analysis, we complement our ALMA dataset with the full suite of available ALMA Band 6 archival data in GOODS-S.

\subsection{GOODS-ALMA}

GOODS-ALMA is a blind 1.1\,mm galaxy survey in the GOODS-S field, centered at $\alpha =\, $3$^{\rm{h}}$32$^{\rm{m}}$30$^{\rm{s}}$, $\delta = -27$\degree48\arcmin00. The survey covers a contiguous area of 72.42 arcmin$^{2}$ with ALMA Band 6 and it is a combination of two observing campaigns carried out at different angular resolutions at a homogenous sensitivity. The high resolution dataset has an average sensitivity of 89.0\,$\mu$Jy beam$^{−1}$ and angular resolution of 0.251"\,$\times$\,0.232" (synthesized beam FWHM). The low resolution dataset has an average sensitivity of 95.2\,$\mu$Jy beam$^{−1}$ and angular resolution of 1.33"\,$\times$\,0.935". The combined mosaic reaches a factor 1.5 deeper sensitivity of 68.4\,$\mu$Jy beam$^{−1}$ at angular resolution of 0.447"\,$\times$\,0.418. The high resolution mosaic was presented in \citet[][GOODS-ALMA 1.0]{Franco_2018}. The low resolution mosaic and the final combination of the two mosaics was presented in \citet[][GOODS-ALMA 2.0]{Gomez_2021}, where we refer the reader for further details about the observations and data processing.

\subsection{ALMA archive}
\label{ALMA_archive}
We downloaded from the archive all ALMA Band-6 projects that cover (i.e., with a primary beam correction better than 0.2) one of our 121 QGs (see Section \ref{Analysis}). This was done using the Python package \texttt{astroquery}. We then calibrated all these projects using the calibration scripts provided in the archive by the ALMA observatory (i.e., the so-called \texttt{scriptForPI.py}). Each QG is covered by at least two ALMA pointings (GOODS-ALMA 1.0 and GOODS-ALMA 2.0) with four ALMA pointings on average. This implies that in the final stacking analysis, a total of 477 ALMA pointings were combined, each with a different depth and spatial resolution. Stacking in the uv-plane with appropriate weightings (see Section \ref{stacking_analysis}) makes it possible to combine all these different pointings.

\begin{figure}
    \centering
    \includegraphics[width = \linewidth]{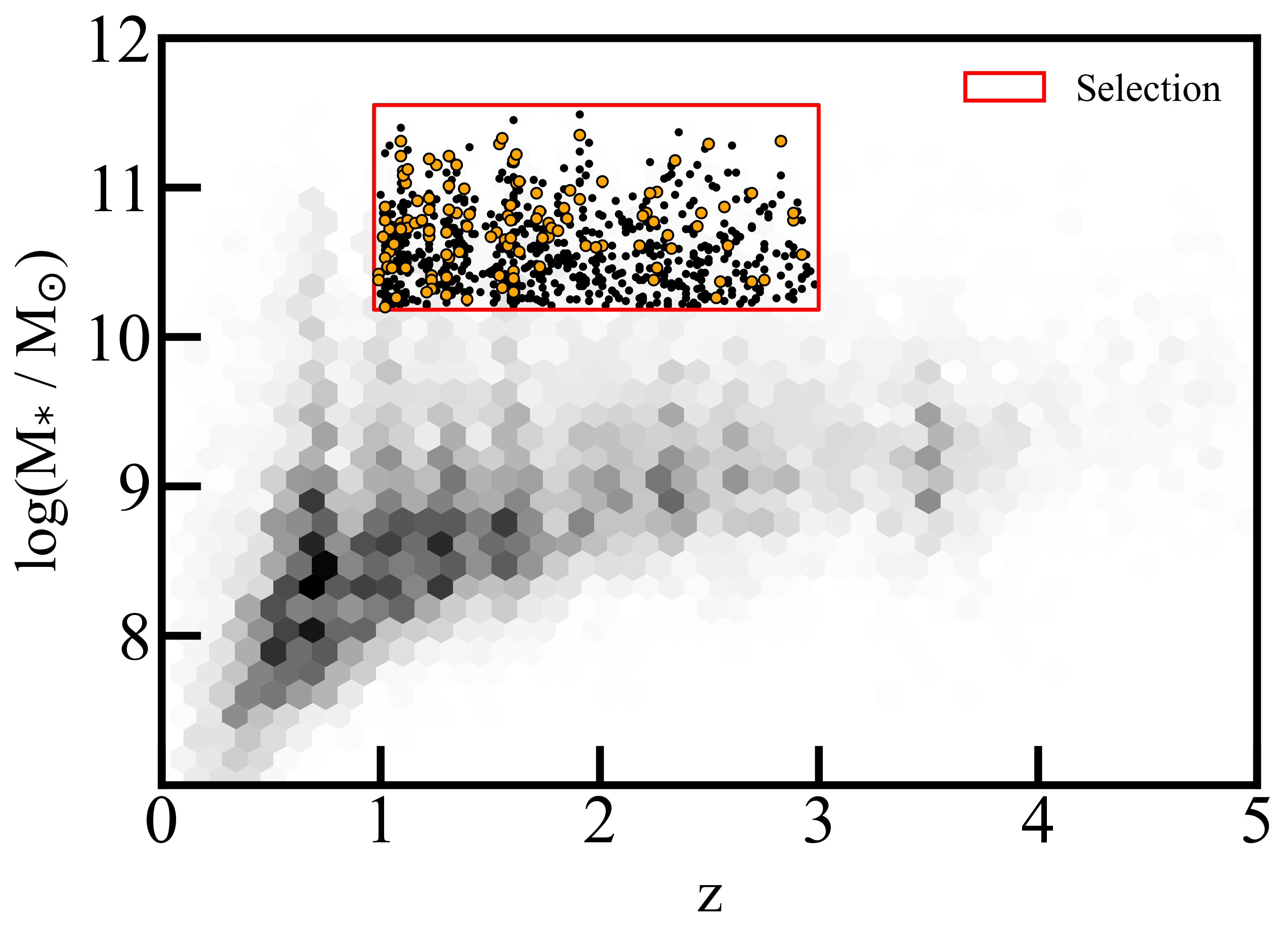}
    \caption{\textbf{Galaxy selection.} Density plot of the used ZFOURGE catalogue in the redshift vs stellar mass plane, consisting of a total 13299 galaxies. The red shaded area represents the region covered by our redshift and stellar mass selection criteria ($1 < z < 3$ and 10.20 $<$ log(\Mstar/\Msol) $<$ 11.50) embedding 852 sources. The orange circles correspond to the QGs that constitute our final selection. }
    \label{fig:galaxy_selection}
\end{figure}

%\begin{figure}
    %\centering
    %\makebox[\linewidth][c]{\includegraphics[width =1.2\linewidth]{Figures/GOODS_ALMA_selection.png}}
    %\caption{Spatial distribution of the sources in our redshift and stellar mass selection criteria. The red points show 435 redshift and mass selected galaxies within the GOODS-ALMA map (depicted in grey with a blue rectangular region highlighting the edges of the map suffering from a poorer sensitivity).}
    %\label{fig:GOODS_ALMA_selection}
%\end{figure}

%%%%%%%%%%%%%%%%%%%%%%%%%%%
%%%%%%%%%%%%%%%
%%%%%%%%%%%%%%%%%%%%%%%%%%%

\section{Sample selection}
\label{Analysis}

In order to select QGs in the GOODS-S field we employ the ZFOURGE catalogue \citep{Straatman_2016}, which contains multi-wavelength ultraviolet (UV) to near-infrared (NIR) photometry of 30911 $K$-band detected sources. The detection image is produced from a combination of  Fourstar/$K_{s}$-band observations with previous $K$-band images in the field \citep{Retzlaff_2010,Hsieh_2012,Fontana_2014}, reaching a 5$\sigma$ detection limit that varies varies between 26.2 and 26.5 mag across the field. The catalogue is complemented with photometric redshifts (photo$-z$) derived with the EAZY fitting code \citep{Brammer_2008} as well as with the fundamental physical properties of the sources (\Mstar, SFR, $A_{\rm V}$, age) inferred by FAST \citep{Kriek_2009}, after adopting the photo$-z$ estimates from EAZY with the stellar population models of \citet{Bruzual_2003}, an exponentially declining star formation history, fixed solar metallicity, and a \citet{Calzetti_2000} dust attenuation law \citep{Straatman_2016}. As a sanity check we also re-run FAST with the same configuration and replace photo$-z$ estimates with spectroscopic redshifts for sources that are included in the most recent spectroscopic catalogue in GOODS-S \citep{Vandels_Garilli}.

We first apply a selection on the ZFOURGE catalogue based on the ($use$=1) flag, removing 17612 photometrically uncertain sources or those that might suffer flux contamination from a nearby star, among other criteria (see \citet{Straatman_2016}, for further information). Since in this work we are interested in typical QGs around cosmic noon, we select 644 sources with $1 < z < 3$ and in the stellar mass range of 10.20 $<$ log(\Mstar/\Msol) $<$ 11.50, as depicted in Fig.~\ref{fig:galaxy_selection}. Then, we further narrow down our sample for sources that fall within the GOODS-ALMA footprint, excluding galaxies lying at the edges of the map, due to their poorer sensitivity. The selected parent sample consist of 435 galaxies that meet our selection criteria.

\begin{figure}
    \centering
     \makebox[\linewidth][c]{\includegraphics[width = 1.15\linewidth]{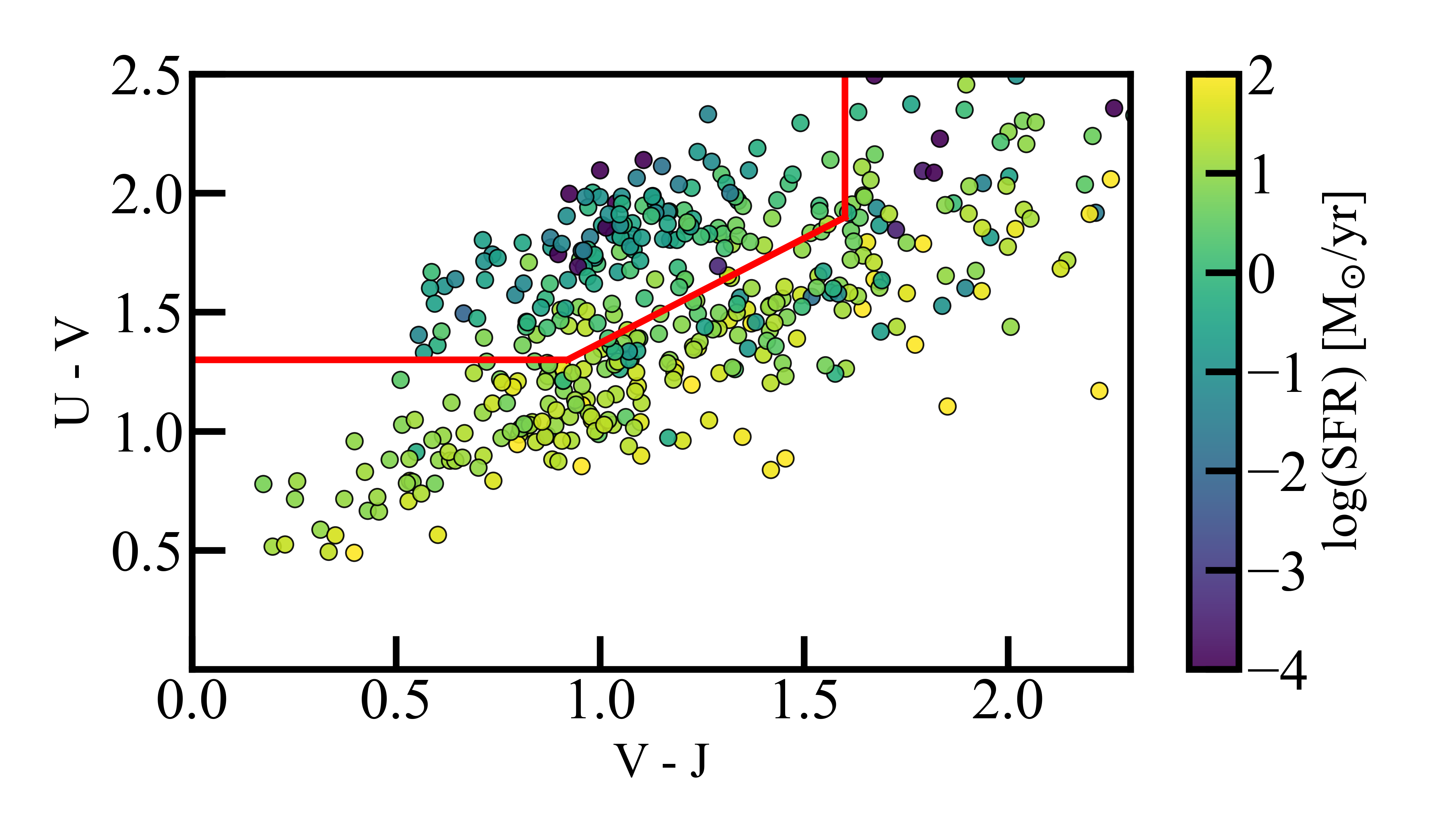}}
    \caption{\textbf{UVJ colour diagram.} Distribution of the parent sample of 435 galaxies that meet our selection criteria in the $U-V$, $V-J$ colour-colour space, colour coded by their log(SFR). The red box represents the quiescent region limits defined in \citet{Schreiber_2015}, which enclose the 140 QGs from which we draw our final sample.}
    \label{fig:UVJ_diagram}
\end{figure}

We then proceed to select QGs from the parent sample using the $UVJ$ criterion \citep{Williams_2009} after inferring the rest-frame colour from the best fit FAST spectral energy distributions (SEDs). For our purposes we adopt the slightly modified colour selection introduced by \citet{Schreiber_2015}: 

\begin{equation}
\begin{cases}
U - V > 1.3,\\V-J < 1.6,\\U-V > 0.88 \times (V-J) + 0.49, 
\end{cases}
\end{equation}
to take into account the different photometric coverage and the  uncertainties in the zero-point corrections. The $UVJ$ colour diagram along with the 140 QGs that meet the adopted colour criteria are shown in Figure \ref{fig:UVJ_diagram}.

In order to validate the quiescent nature of the selected galaxies we also perform some additional tests. First, we compare the inferred SFR of the galaxies in our sample to that of the MS galaxies at the corresponding redshift and stellar mass, quantified as $\delta{\rm MS} = \rm{SFR/SFR_{MS}}$, using the MS prescription of \citet{Schreiber_2015}. Adopting a $\delta{\rm MS} $<1/5 cut we find that our entire sample is also quiescent according to their SFR levels. In the process, we perform a quality check by visually inspecting the best fit SEDs. We also confirm that our selected galaxies are undetected (down to 2$\sigma$) in the $Herschel$/PACS and SPIRE bands, adopting the GOODS-$Herschel$ catalogues presented in \citet{Elbaz_2011}, as well as undetected in the 1.1\,mm GOODS-ALMA 2.0 map \citep{Gomez_2021}. However, we choose to include galaxies that are detected at MIPS 24\,$\mu$m (but not in PACS/SPIRE/ALMA), to avoid biasing our sample against QGs with strong mid-IR emission originating from evolved old stellar populations or intermediate AGN activity \citep{Fumagalli_2014}. This decision does not affect the recovered signal in our stacking analysis. Finally, to avoid contamination from neighbouring sources in our stack, we remove QGs that lie within a distance of 3'' from a GOODS-ALMA detected source. The final sample consist of 121 QGs, with a median $z$ of 1.53$\,\pm\,$0.03 and a median \Mstar\ of (5.5$\,\pm\,$0.2)$\;\times\;$10$^{10}$\,\Msol.

%%%%%%%%%%%%%%%%%%%%%%%%
%%%%%%%%%%%%%
%%%%%%%%%%%%%%%%%%%%%%%%

\section{Stacking analysis}
\label{stacking_analysis}

We perform a stacking analysis of the final sample of 121 QGs at hand  in the $uv$-plane (see methodology in \citet{Gomez_2021} and \citet{Wang_2022}), which has been shown to improve the signal-to-noise ratio (S/N) of the stacked image compared to a stacking analysis performed in the image plane \citep{Lindroos_2015} and most importantly, allows to stack datasets of different angular resolution. In particular, we apply mean $uv$-stacking to the QGs sample concatenating the visibilities from the GOODS-ALMA observations. The main advantage of the GOODS-ALMA map is the homogeneous coverage of the field that results in a uniform contribution of the stacked sources in the final image.

Subsequent aperture photometry at the phase center yields a 3$\sigma$ upper limit flux of  47.3 $\mu$Jy. While the measurement is below a formal detection limit ($ > 3\sigma$), the curve of growth of the emission at phase center follows closely that of the PSF from the stack, hinting that the recovered signal could originate from a real source, which could be detected if deeper observations were available. This curve of growth could also be caused by a noise peak located right at the phase center. However the probability of this scenario is extremely low (< 0.05\%).
 
\begin{figure}[t]
    \centering
    \includegraphics[width = \linewidth]{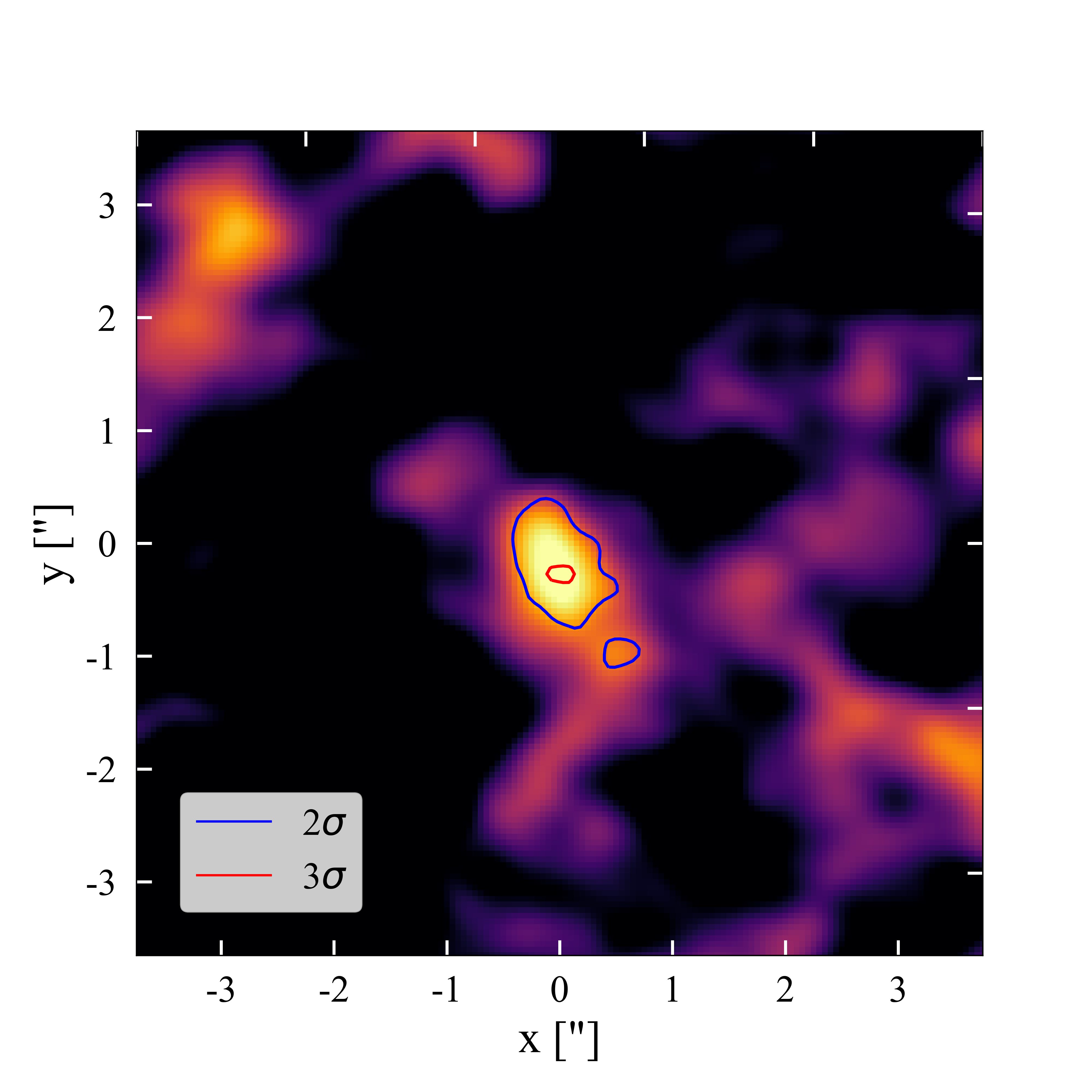}
    \caption{\textbf{Stacked image.} 7"$\times$7" image cutout of the ALMA Band 6 archival $uv$-stack centered at the phase center. The blue and red contours represent the 2$\sigma$ and 3$\sigma$ flux emission levels respectively.}
    \label{fig:UV_stack}
\end{figure}

In order to improve the sensitivity of the stack, we decide to concatenate visibilities from the entire ALMA archival observations in Band 6 in the GOODS-S field, as described in Sect.~\ref{ALMA_archive} (see Fig. \ref{fig:UV_stack}). Once more, we perform aperture photometry at the phase center. We apply a minor recentering of the aperture position according to the uncertainty in the astrometric accuracy of the ALMA pointings, which according to the ALMA technical handbook is calculated as:

\begin{equation}
    {\Delta}S = \frac{\mathrm{Beam_{FWHP}}}{0.9\;{\times}\;\mathrm{S/N}}.
\end{equation}

\noindent Approximating the beam size to $\mathrm{Beam_{FWHP}}\sim$ 0.5", we estimate an astrometric uncertainty $\Delta$S$\sim$ 0.18". We then relocate the aperture center position by $\Delta{\alpha}$= -0.05" and $\Delta{\delta}$ = -0.10", these being the changes in right ascension and declination respectively, in order to optimize the S/N.

\begin{figure}[t]
    \centering
        \makebox[\linewidth][c]{\includegraphics[width = 1.15\linewidth]{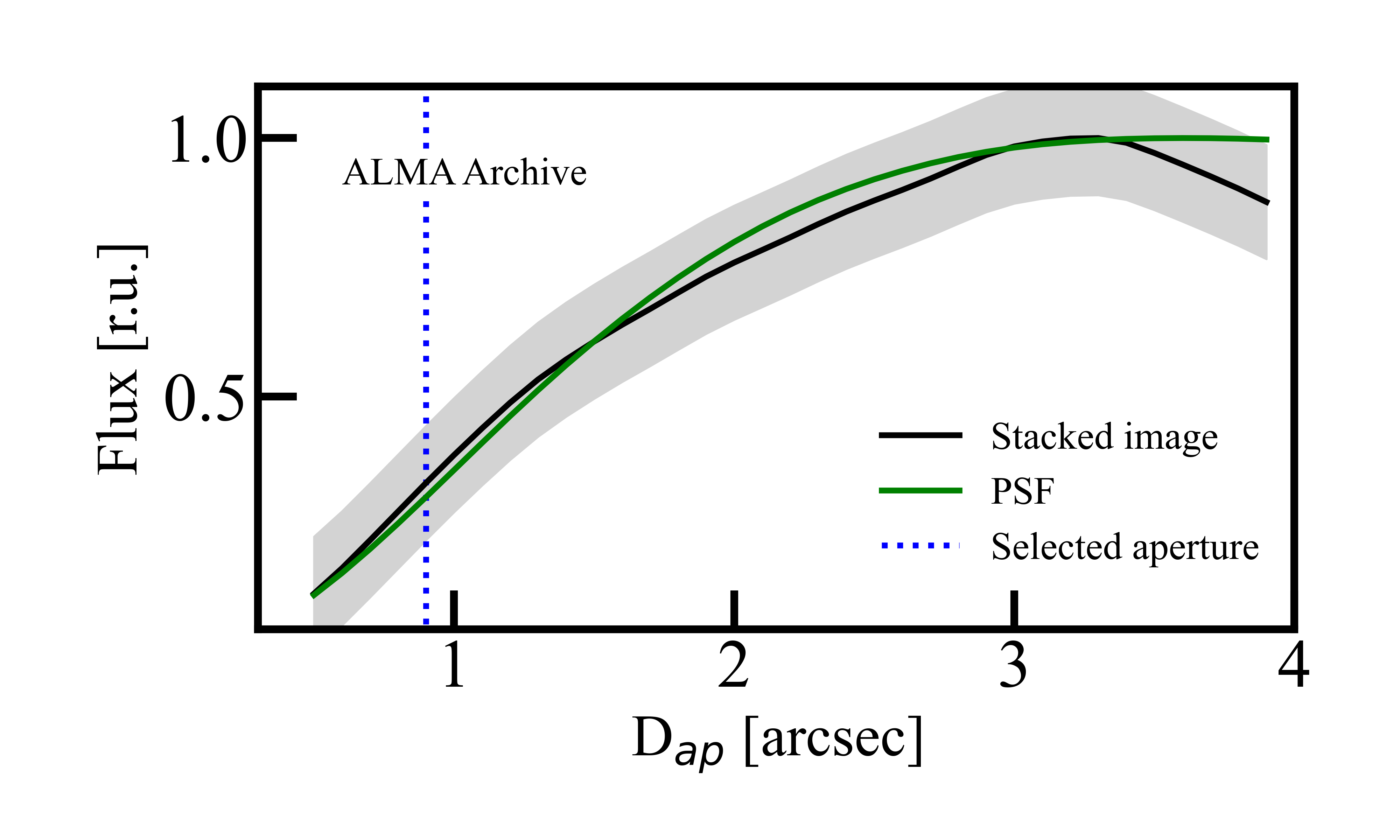}}
    
    \caption{\textbf{Image analysis.} Curve of growth at the phase center of the ALMA archival $uv$-stacked image (black), with its associated uncertainty shown as grey shaded area, and of the PSF (green). The blue dotted line depicts the selected aperture diameter of 0.9".}
    \label{fig:PSF_aper}
\end{figure}

In order to estimate the uncertainty of the measured flux density, we randomly place 3000 empty apertures around the phase center and adopt the standard deviation of their measured flux density distribution as the error. The diameter of the placed apertures is the same as the one we selected to make the source flux density measurement. We select an aperture size that maximizes the S/N of the recovered flux and is also consistent with the expected dust component size of QGs at $z \sim$ 1.5. Recent results indicate that the dust continuum emission is expected to be smaller than the stellar emission \citep{Whitaker_2021}. Therefore, using a prior conservative estimate of the stellar component size for QGs at $z \sim$ 1.5 and log(\Mstar/\Msol) = 10.7, i.e. $R_{\mathrm{eff}} \sim 0.2"$ \citep{VanderWel_2014}, we define an aperture with $D_{\rm{ap}}$ = 0.9", allowing to capture the entire dust emission of the stack. The flux is then corrected by the appropriate aperture correction to account for the flux losses outside the aperture, calculated by dividing the flux within the aperture of $D_{\rm{ap}}$ = 0.9" by the flux enclosed in the synthesized dirty beam of the stack (Beam$_{\mathrm{FWHM}}\approx$0.5") within the same aperture (normalised to its maximum value, see Fig.\ref{fig:PSF_aper}), yielding a correction factor of 3.26.

The described methodology finally yields a flux measurement of 11.70 $\pm$ 3.59\,$\mu$Jy with S/N = 3.25, value that we use for the rest of the study. Concerning the physical properties of the stacked QGs sample, the heterogeneity in the integration time of each galaxy from the ALMA archive weighs differently each individual source. The analysis of the weight distribution of the stack shows three distinct objects with weights an order of magnitude higher than the mean value. These three galaxies bias the entire stack to a misleading higher \Mstar\ and a lower $z$ when compared to the remaining sample. Therefore, we remove these outliers and perform the same stacking analysis. The resulting S/N is hardly affected (only from the third significant digit on). The weighted average physical properties of the quiescent sample vary slightly from ${\langle}z{\rangle}$ = 1.62$\,\pm\,$0.04 and ${\langle}$\Mstar$\rangle$ = (6.55$\,\pm\,$0.39$)\;\times\;$10$^{10}$ \Msol, associated to the homogeneously weighted GOODS-ALMA stack, to ${\langle}z{\rangle}$ = 1.47$\,\pm\,$0.03 and ${\langle}$\Mstar$\rangle$ = (5.00$\,\pm\,$0.26$)\;\times\;$10$^{10}$ \Msol, associated to this latter ALMA archival stack.

%%%%%%%%%%%%%%%%%%%%%%%%%
%%%%%%%%%%%
%%%%%%%%%%%%%%%%%%%%%%%%%

\section{Gas fraction}
\label{gas_fraction}

The observed frame 1.1\,mm emission from galaxies at $z \sim 1-3$ samples the Rayleigh-Jeans (R-J) tail of dust emission. In order to obtain a dust mass estimate from the measured stacked flux density we utilise the QG SED model shown in \Magdis, which is calibrated to make valid predictions for the average QG population. For a more detailed explanation of the utilised data and techniques to construct this QG SED model we refer to \Magdis. This empirical model is characterised by a dust temperature ($T_{\mathrm{dust}}$) of 20\,K. The observed weak evolution in the $T_{\mathrm{dust}}$ of massive QG makes this average temperature a robust estimate for their ISM. This has lately been found consistent by simulations in \citet{Cochrane_2022}, who has used the FIRE (Feedback In Realistic Environments) project \citep{Hopkins_2014} to study the dependence of dust mass on FIR flux and $T_{\mathrm{dust}}$.

Fitting a unique model for a stack ensemble covering a large redshift range intrinsically assumes that the FIR flux density at the observed 1.1\,mm does not evolve significantly with cosmic time. As shown in \Magdis, redshift has a negligible effect in the flux density in ALMA Band 6 at $1\,<\,z\,<\,3$. This is attributed to the negative K-correction, i.e., the counter acting between the flux dimming due to increasing cosmic distance and the flux boosting due to the redshifting of the galaxy light together with the negative slope of the R-J tail.

\begin{figure*}
    \centering
    \includegraphics[width = 0.78\linewidth,pagecenter]{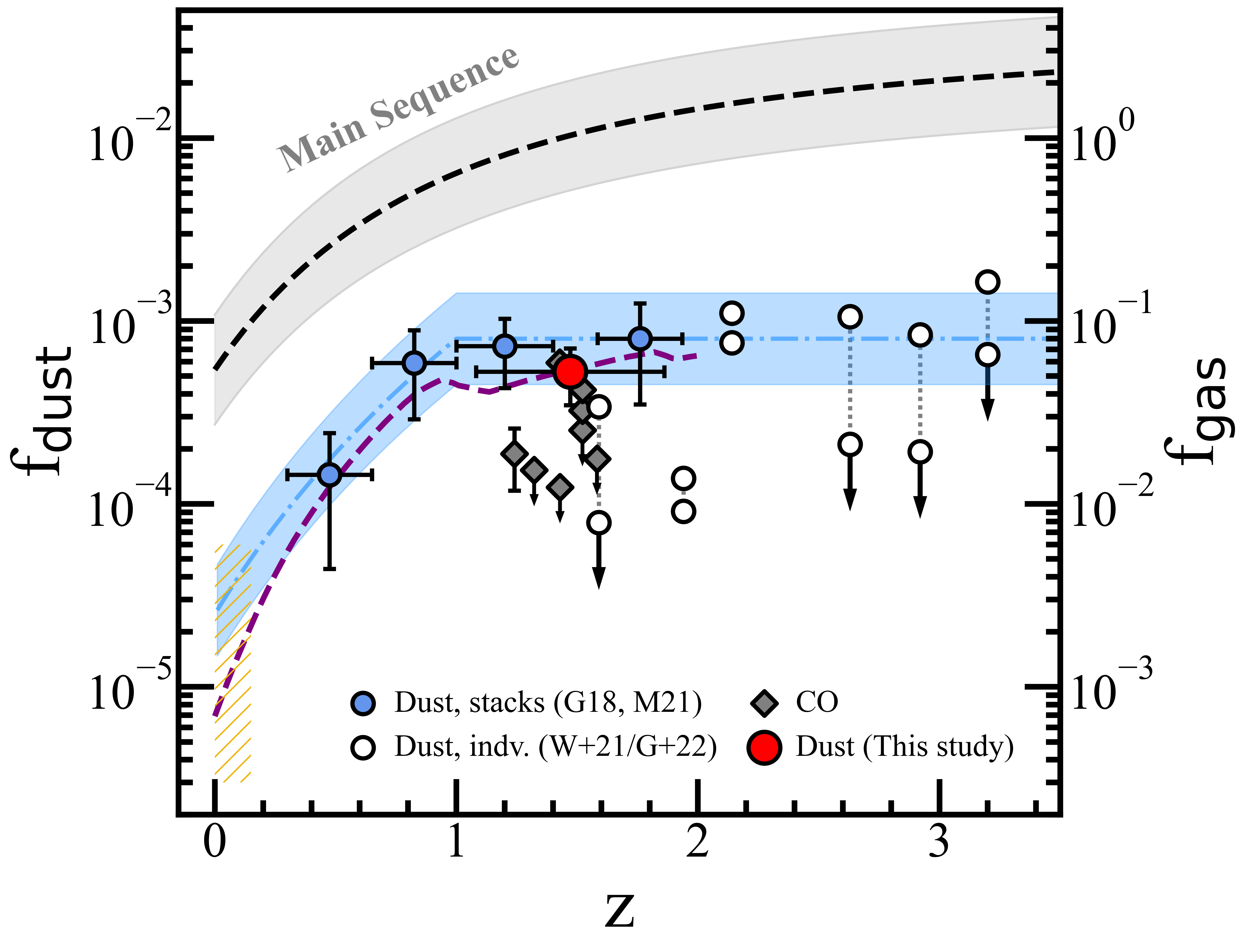}
    \\
    \includegraphics[width = 0.78\linewidth,pagecenter=18cm]{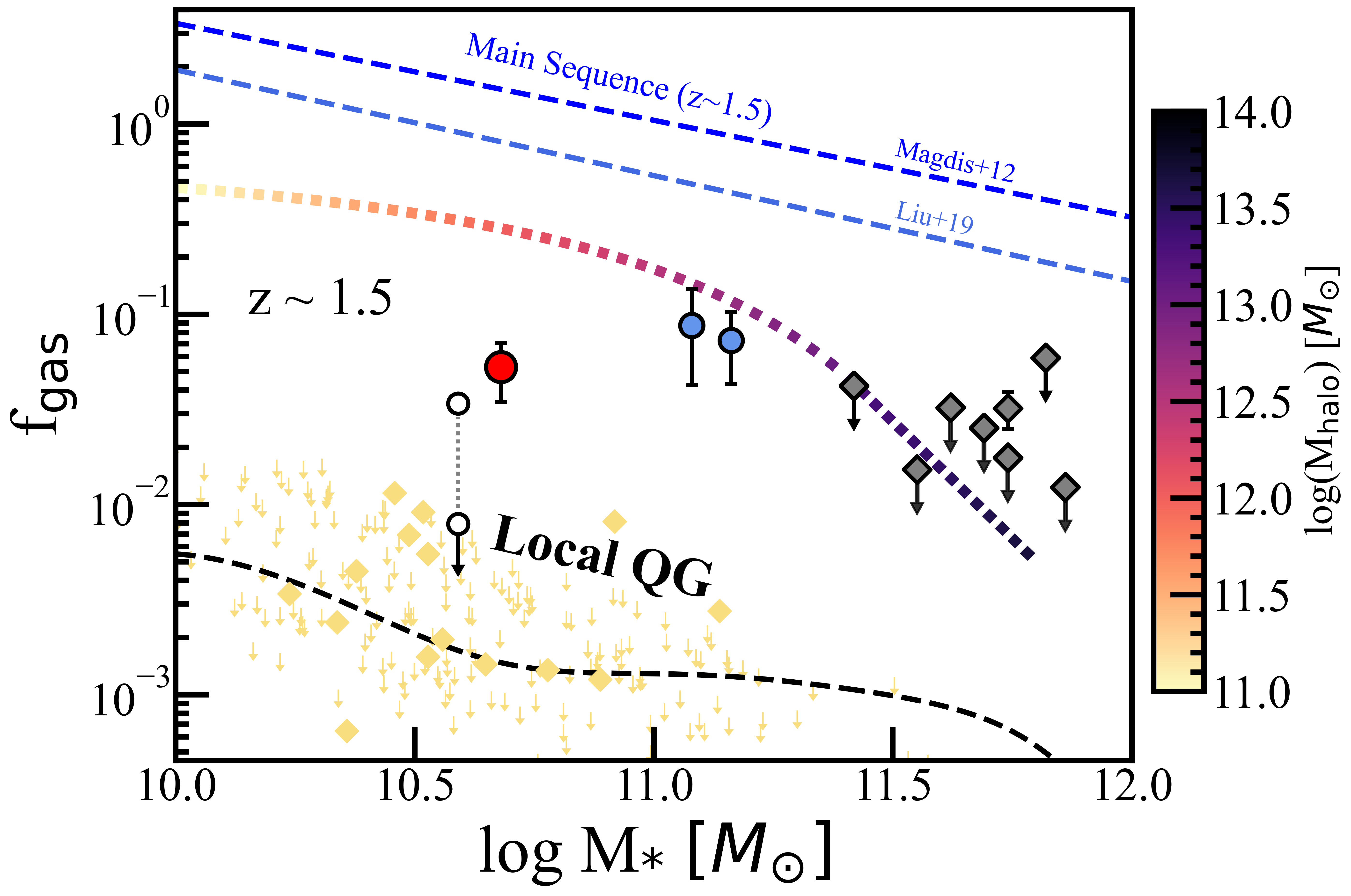}

    \caption{\textbf{Gas and dust fractions of QGs} \textit{Top}: Selection of \fdust\ and \fgas\ measurements as a function of  redshift for QGs. Circles correspond to dust derived gas fractions: this work and  previous stacks studies (\Gobat\ and \Magdis) are shown in red and blue, respectively. The white circles show two different estimates (connected by a grey dotted line) for a sample of individually observed lensed galaxies. The lower values correspond to those presented in \citet{Whitaker_2021} and the upper values show the new estimates provided in \citet{Gobat_2022}. The grey diamonds represent CO derived \fgas\ estimates \citep{Sargent_2015,Bezanson_2019,Williams_2021}. The red dashed area embeds \fgas\ measurements of local QGs obtained for the ATLAS3D sample \citep{Young_2011,Cappellari_2013,Davis_2014}. The blue shaded area and the purple dashed line represent the best fit to the \Magdis\ data and the \citet{Gobat_2020} model respectively. For reference, we add the \fgas\ evolution of main sequence galaxies according to \citet{Liu+19}. \textit{Bottom}: Dust and gas fraction as a function of stellar mass for measurements at $z \sim 1.5$. The symbols are the same as in the top panel. The dotted line shows the \fgas\ prediction according to the \citet{Dave_2012} galaxy evolution models, color coded as a function of $M_{\mathrm{halo}}$. For reference, we add the \fgas\,-\,\Mstar\ trend measured by \citet{Magdis_2012,Liu+19}. The light yellow scattered diamonds and arrows mark the \fgas\ detections and upper limits for local QGs with the corresponding best fit plotted as a black dashed line.}
    \label{fig:f_gas_plot}
\end{figure*}

For the calculation of the $M_{\mathrm{dust}}$ of our compiled stacked ensemble we bring our flux density measurement to the rest frame and compare it to that of the model, resulting in an scaling factor $N$.

\begin{equation}
    \mathrm{N} = \frac{f_{\nu}}{f^{0}_{\nu}}, 
\end{equation}

\noindent where $f_{\nu}$ and $f^{0}_{\nu}$ are the measured flux density and the normalised SED model flux density at observed frame 1.1\,mm, respectively. Since the cold gas mass scales linearly with the \Mdust, and the latter can be linearly scaled from the original SED, we can calculate the corresponding gas fraction as:

\begin{equation}
    f_{\mathrm{gas}} = \frac{\mathrm{N}\;{\times}\;\mathrm{GDR}(Z){\;\times}\;M^{0}_{\mathrm{dust}}}{M_{*}}, 
\end{equation}

\noindent where $M^{0}_{\rm{dust}}$ is the dust mass of the normalised SED model and $\rm{GDR}(Z)$ is the gas-to-dust ratio, which corresponds to $\rm{GDR}(Z_{\odot})$ = 92 when adopting a universal solar metallicity \citep{Leroy_2012,Magdis_2012}. Following this prescription we obtain an estimate for \Mdust = (2.82\,$\pm$\,0.86) $\times$\,10$^{7}$\,\Msol , which we convert to an \Mgas\ = (2.63\,$\pm$\,0.79) $\times$\,10$^{9}$\,\Msol. We then calculate the corresponding \fdust\ = 0.05\,$\pm$\,0.02\,\% and \fgas\ = 5.3\,$\pm$\,1.8\,\% (see Fig.~\ref{fig:f_gas_plot}). For completeness, we also calculated \Mgas\ using the monochromatic gas mass estimate method presented in \citet{Scoville_2017}, which yields a \Mgas\ = (1.8\,$\pm$\,0.56) $\times$\,10$^{9}$\,\Msol\ and a corresponding \fgas\ = 3.5 \,$\pm$\,1.19\,\%. This decrease is fully consistent with the higher dust temperature adopted in this method to convert the FIR flux density to gas mass. While estimating \Mgas\ through \Mdust, i.e., using a metallicity-dependent GDR, introduces an additional uncertainty in our analysis, the adopted value yields the most conservative estimate. We note that recent work by \citet{Morishita_2022} on a gravitationally lensed QG has shown that the normally assumed $\rm{GDR}(Z_{\odot})$ could underestimate the actual gas mass by a factor of $\times$\,1.6, while the SIMBA cosmological simulation indicates a range of 4 orders of magnitude in the GDR of QGs \citep{Whitaker_2021b}.

In Figure \ref{fig:f_gas_plot} we show the derived $f_{\mathrm{dust}}$ and \fgas\ as a function of redshift along with a collection of estimates from the literature including local QGs \citep{Young_2011,Cappellari_2013,Boselli_2014,Davis_2014}, high$-z$ stacked ensembles (\Gobat, \Magdis) and dust continuum observations of high$-z$ QGs \citep{Whitaker_2021}. For consistency, we derive \Mdust$,\,$\Mgas$,\,$\fgas\ using the same methodology as that described in Sect.~\ref{stacking_analysis} to the dust continuum measurements of \citet{Whitaker_2021}. This results in a mean 30$\%$ increase in \fgas\ compared to their presented values. Nevertheless, this only reinforces the conclusions presented below and in Sect.~\ref{Discussion}. The depleted gas reservoirs of local QG with \fgas\ $<$ 1\,\% is in direct contrast with the increasing \fgas\ values of the stack ensembles estimate from $z$ = 0 to $z\,\sim$\,1. The steep increase is followed by a flat evolution towards higher redshift. This behaviour is partially mirrored by the evolution of \fgas\ in MS SFGs, approximately 1.5\,dex higher. On the other hand, the individual observations of QGs show, on average, a much weaker evolution in the gas budget of QGs with redshift, predominantly yielding upper limits in \fgas\ that are in tension with the estimates of stacks. 

Our measurement appears to be consistent with previous stacking analyses (\Gobat, \Magdis), indicating a non negligible amount of gas reservoir in the average population of QGs at $z\sim1.5$ and an increase by a factor of $\times$\,10 in \fgas\ with respect to their local counterparts. We recall that the previous  studies were based on the stacking of approximately 1500 sources in  $Herschel$/SPIRE 250, 350, 500$\;\mu$m, JCMT/SCUBA-2 850$\;\mu$m, ASTE/AzTEC 1.1$\;$mm maps with a resolution that ranges from 15" to 36", yielding a 0.12 mJy upper limit (3$\sigma$) at 1.1$\;$mm. The sub-arcsecond resolution of the ALMA observations considered here, yielding a clear detection at 1.1mm, seem to suggest that the reported tension between stacked results and the studies of individual QGs reported in the literature, is not, fully at least, originating from the poorer resolution of the former.

\section{Discussion}
\label{Discussion}
The analysis described in the previous section yields a gas fraction of \fgas\ = 5.3\,$\pm$\,1.8\% for the average population of quenched galaxies at $z\,\sim\,1.5$. This estimate is consistent with the values derived in the stacking analysis of \Gobat\ and \Magdis, indicating that high-$z$ galaxies, on average quench with non-negligible amounts of gas still being present.

\subsection{Gas fraction as a function of stellar mass}
We also want to explore the \fgas\ - \Mstar\ relation for high-$z$ QGs. For the star forming galaxy population an anticorrelation between these two parameters has been found at both the local universe \citep{Cicone_2017,Saintonge_2017} and high-$z$ \citep{Sargent_2014,Liu+19}. The same has also been confirmed for local QGs \citep{Boselli_2014,Saintonge_2022}. This trend yet remains to be investigated for high-$z$ QGs due to the narrow dynamical range in \Mstar\ of the data obtained up to date. So far,  the scarce \fgas\ measurements of distant QGs have essentially focused on the very massive population (log(\Mstar/\Msol) $>$ 11.2) \citep{Williams_2021} or gravitationally lensed systems \citep{Whitaker_2021}. Our stacking analysis, capturing the intermediate mass population of QGs, allows us to probe this trend.

In Fig.~\ref{fig:f_gas_plot}, we show for the first time the \fgas\ as a function of \Mstar\ at $z\,\sim\,1.5$. As a reference, we also plot the well established trend  for MS SFGs at the same redshift as well as for local QGs. First, we observe that all QGs measurements lie well below the MS (\fgas$\lesssim 15$\,\%). More importantly, all the upper limits, consistent with the gas depletion picture, are clustered at the massive end of the plot. Indeed, spectroscopic studies and imaging of individual high$-z$ QGs have so far been heavily biased towards the most extreme and massive systems (log \Mstar\ $>$ 11.3). Conversely, the stacks, supportive of a gas retention scenario, reflect an average gas fraction that is  representative of the more typical, moderately massive QGs at $z\,\sim$\,1.5. This suggests that the apparent tension between the studies of stack ensembles and of individual sources can be attributed to selection effects, with more massive QGs having experienced a larger depletion of their gas mass reservoir at the same redshift. 

Nevertheless, the large uncertainties of our estimates together with the lack of data at the low mass end of high-$z$ QG population prevent us from fully uncovering the dependence of \fgas\ on \Mstar. Indeed, whether the \fgas\ remains flat or increases at lower stellar masses, as is the case for SFGs or local QGs, cannot be constrained with the current data. 

Finally, it is worth considering some theoretical predictions regarding the dependence of \fgas\ on \Mstar. For that, we consider the analytic model presented by \citet{Dave_2012} which is based on the assumption that galaxy growth is regulated by an equilibrium between star formation, gas inflows and outflows. We note that the model can only reproduce QGs at $z \sim $1.5 that reside in the most massive dark matter halos (log($M_{\mathrm{halo}}$/\Msol) $>$ 13.5), and thus the comparison to our data is only valid in the high mass end. For log(\Mstar/\Msol) $>$ 11.2, the model predicts a sharp decline in the \fgas\ of QGs, consistent with the observed upper limits (Fig.~\ref{fig:f_gas_plot}), further supporting the scenario where more massive QGs, that reside in more massive haloes, exhibit lower \fgas\ that those of lower \Mstar.

\subsection{Caveats}

It is acknowledged that the stacking analysis and the subsequent derivation of \Mdust\ and \Mgas\ from  dust continuum observations come with a suite of inherent uncertainties, caveats and implicit assumptions. As we discussed and addressed in the previous sections these include the selection of QGs based on the UVJ diagram, the assumption of homogeneity in the properties of the population (inherent to any stacking technique) and the adopted methodology to convert the observed flux densities to physical parameters (e.g. adopted template, dust temperature, GDR). On top of these, the spatial extent of stacked sources should also be considered. 

\begin{figure}[t]
    \centering
    \includegraphics[width = \linewidth]{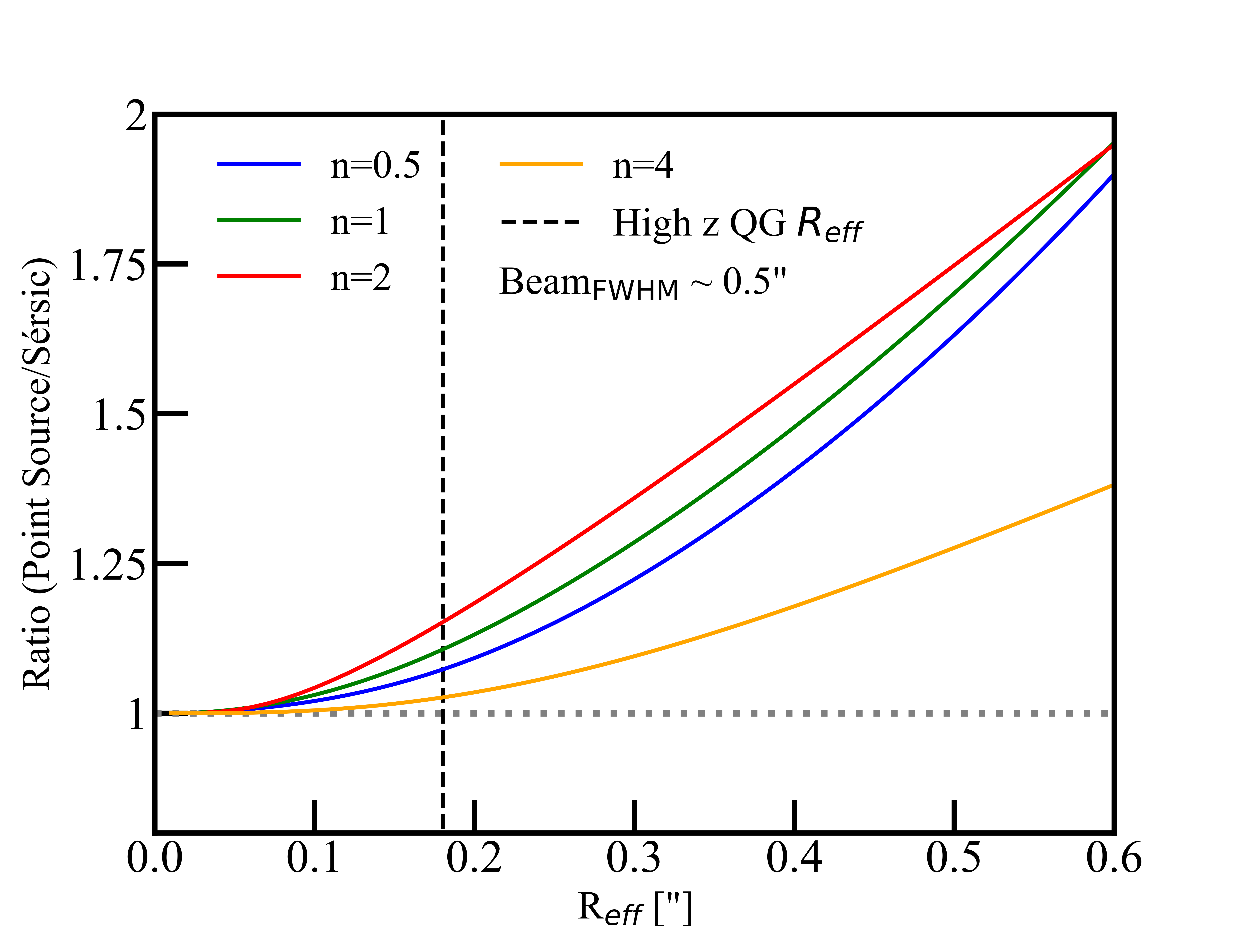}
    \caption{\textbf{Size assumption effect.} Flux correction factor ratio  between the assumed point source emission and a modelled galaxy profile as a function of effective radius. Each colored line shows the ratios for profiles with different S\'ersic indexes. The black vertical dashed line represents a size reference of $R_{\mathrm{eff}}$ based on the stellar component size of QGs at $z \sim 1.5$ with log(\Mstar/\Msol) $=$ 10.7 \citep{VanderWel_2014}.}
    \label{fig:sersic index}
\end{figure}

The adopted aperture correction applied to measured flux density assumes that the object at the phase center is either a point source or marginally resolved. We test the robustness of this assumption by estimating how much flux would not be measured should the object not be a point source. To do so we convolve a number of S\'ersic profiles \citep{Sersic_1963} with the ALMA-archive synthesized dirty beam and study how the flux changes as a function of the source size. In Figure \ref{fig:sersic index} we show the ratio of missed flux between a point source and galaxy profiles with varying S\'ersic indexes $n$ as a function of the galaxy effective radius $R_{\mathrm{eff}}$. Measurements of the stellar sizes of QGs at $z\,\sim$\,1.5 with log(\Mstar/\Msol) $=$ 10.7, as traced by the optical wavelengths, yields a $R_{\rm{eff}}\,\sim$\,0.18" \citep{VanderWel_2014}, which we use as a reference. We notice that, for a typical QGs $n = 4$, using a conservative $R_{\rm{eff}}\,\sim$\,0.18", the flux would be underestimated by $\sim$ 5\%, yielding a \fgas\ $\sim$ 6\%, and thus, reinforcing the gas retention scenario and subsequent conclusions in Sect.~\ref{gas_fraction} and in the discussion.

\section{Conclusions}

We conducted a 1.1\,mm stacking analysis of a carefully selected sample of $1 < z < 3$ QGs, using both the GOODS-ALMA survey and the full ensemble of ALMA archival data in the GOODS-S field. We recover at the phase center of the image a signal of $S_{\rm 1.1mm}$ = 11.70\,$\pm$\,3.59 $\mu$Jy arising from the stacked ensemble of QGs with weighted mean values of ${\langle}z{\rangle}$ = 1.47  and 
log${\langle}$\Mstar$\rangle$ = 10.70\,\Msol. Using the empirical templates of \Magdis, we estimated an average log(\Mdust/\Msol) = 7.47\,$\pm$\,0.13 and log(\Mgas/\Msol) = 9.42\,$\pm$\,0.14 that correspond to \fdust\ = 0.05\,$\pm$\,0.02\,\%  and \fgas\ = 5.3\,$\pm$\,1.8\,\%. With these estimates along with with previous stacking results and studies of individual high-$z$ QGs we reach the following conclusions: 

\begin{enumerate}
    \item The average population of QGs at $z = 1-3$ appears to retain a non-negligible amount of gas, with an average \fgas\ = 5.3\,$\pm$\,1.8\,\%. This is in agreement with previous stacking results but considerably larger compared to the values, or upper limits inferred by studies of individual high$-z$ QGs. 
    
    \item While still poorly constrained due to the scarcity of data in the low mass regime, the \fgas\ of QGs at $\langle z \rangle$ = 1.5 appears to remain flat up to log(\Mstar/\Msol)\,$\sim$\,11, and then rapidly drop towards the largest stellar masses. This could alleviate the tension between the stacking results and the studies of individual sources that so far have predominantly drawn targets from the very massive end of the high$-z$ population of QGs.

\end{enumerate}

  The sub-arcsecond resolution of the ALMA observations exploited in this study has been instrumental into overcoming the main caveat of the previous stacking attempts of high$-z$ QGs, i.e. the uncertainties introduced by the coarse resolution of \textit{Herschel} and other ground based sub(mm) facilities. However, the debate between gas depletion versus gas retention in high$-z$ QGs and the physical mechanisms involved in the truncation of the star formation of galaxies is still on and  requires deeper continuum observations or resorting to alternative gas mass tracers \citep[e.g. {[CII];}][]{Zanella_2018}.

\begin{acknowledgements}
GEM and DBS acknowledge financial support from the Villum Young Investigator grant 37440 and 13160 and the Cosmic Dawn Center (DAWN), funded by the Danish National Research Foundation under grant No. 140 PD. We also thank the anonymous referee for their instructive comments. 
\end{acknowledgements}

\bibliographystyle{aa}
\bibliography{references.bib}

\end{document}